\documentclass[11pt]{iopart}
 
\usepackage{graphicx, color, amssymb}
\usepackage{multirow}

\begin{document}

\title[Nanospintronic devices]{Fabrication and characterisation of nanospintronic devices}

\author{J Samm, J Gramich, A Baumgartner, M Weiss and C Sch\"onenberger}

\address{
Institute of Physics, University of Basel, Klingelbergstrasse 82, 4056 Basel, Switzerland\\}

\ead{andreas.baumgartner@unibas.ch}

\date{\today}

\begin{abstract}

We report an improved fabrication scheme for carbon based nanospintronic devices and demonstrate the necessity for a careful data analysis to investigate the fundamental physical mechanisms leading to magnetoresistance. The processing with a low-density polymer and an optimised recipe allows us to improve the electrical, magnetic and structural quality of ferromagnetic Permalloy contacts on lateral carbon nanotube (CNT) quantum dot spin valve devices, with comparable results for thermal and sputter deposition of the material. We show that spintronic nanostructures require an extended data analysis, since the magnetisation can affect all characteristic parameters of the conductance features and lead to seemingly anomalous spin transport. In addition, we report measurements on CNT quantum dot spin valves that seem not to be compatible with the orthodox theories for spin transport in such structures.

\end{abstract}

\pacs{85.75.-d, 72.25.-b, 75.47.-m}


\maketitle



\section{Introduction}

Merging the exquisite tunability of electronic nanostructures with ferromagnetic materials in nanospintronic devices bears great potential for applications and fundamental investigations. 
Electronic devices using the electron spin in magnetic field sensing are very successful, for example in hard disks of computers. However, to use the electron spin directly, for example in a spin-transistor \cite{Datta_Das_APL56_1990} or as quantum bits \cite{Loss_DiVincenzo_PRA58_1998}, it is necessary to fabricate nanostructures with the required long coherence times and electrical tunability.
Carbon nanotubes (CNTs) and graphene are in principle ideally suited for spintronic devices due to the large intrinsic coherence times, tunable electron density and large maximum current densities. Early electrically tunable spin valves on carbon nanotubes \cite{Sahoo_Kontos_Schoenenberger_NaturePhys1_2005, Man_Morpurgo_PRB73_2006}, or nonlocal spin-accumulation experiments on graphene \cite{Tombros_VanWees_Nature448_2007} demonstrate the great potential of carbon based nanostructures. To obtain an electrically tunable spin signal, one strategy is to fabricate a nanostructure with a gate-tunable conductance, e.g. a quantum dot (QD) \cite{Sahoo_Kontos_Schoenenberger_NaturePhys1_2005, Aurich_Baumgartner_APL97_2010}. However, this electrical tunability introduces additional complexity to the data analysis, since the signal now depends on the position, amplitude and broadening of a conductance feature, which all can vary with the magnetisations of the contacts, as will be discussed in the results sections of this paper. While in nonlocal measurements the spin signal can in principle be separated from the charge signal, this is difficult in two-terminal QD devices because both signals are detected at the same contacts. Despite considerable efforts, spin injection and detection in QD spin valves are not yet reproducible enough for more complex experiments or applications, e.g. as detectors of electron spin entanglement \cite{Kawabata_JPhysSocJap_2001, Klobus_Baumgartner_Martinek_ArXiv_1310.5640v2}.

This lack of reproducibility can have several reasons. The most fundamental spin transport device is a spin valve with two ferromagnetic (F) contacts to a non-magnetic material in-between. In Fig.~1a such a device is shown schematically with a carbon nanotube (CNT) between the F-contacts. Ideally, the contacts are either magnetized parallel or anti-parallel to each other, adjustable by an external magnetic field. The normalized difference between the electrical conductance (or resistance) of these two configurations is called magnetoresistance (MR). Here we address two more technical problems. The first is limitations in the device design. Compared to other carbon based nanoscale devices with normal metal \cite{Chen_Avouris_Nanoletters_2005} or superconducting leads \cite{Schindele_Baumgartner_Schoenenberger_PRL109_2012}, the contact material has to be chosen from a very limited range of readily available and processable magnetic metals, which limits the optimization of the contacts. In addition, most ferromagnetic materials form oxides when exposed to air, which diminishes the electrical contact yield. To obtain low-ohmic contacts with non-magnetic materials, one often chooses large contact areas, which, however, is in conflict with using narrow contact geometries to control the shape anisotropy and thus the magnetic field at which the magnetisation is reversed (switching field) \cite{Aurich_Baumgartner_APL97_2010}. Even the thickness of the deposited material is limited to avoid the formation of vertical, more complex magnetic domains. No adhesion or contact layer can be used because the equilibrium spin polarization decays very rapidly  in non-magnetic metals (on the scale of the exchange interaction, typically $<1\,$nm). Our choice of contact material is the well-studied Ni$_{80}$/Fe$_{20}$ alloy Permalloy (Py), for which one can obtain single-domain contacts and control over the magnetic easy axis by the shape of the contacts \cite{Aurich_Baumgartner_APL97_2010}. We demonstrate that the same electrical and magnetic characteristics for sub-micrometer scale Py contacts can be obtained by sputter deposition and for thermal evaporation. This opens up the large field of magnetic multi-layer structures to be used in carbon based nano spintronic device fabrication, e.g. anti-ferromagnetic exchange-bias layers \cite{Nogues_PhysRep_2005}.

The second technical problem we propose a solution to is resist residues and the unwanted formation of Py nanoparticles near and on top of the device, which can strongly alter the device characteristics. In nanospintronics the interface area between the ferromagnetic contact and the non-magnetic structure, for example a CNT, is usually very small, which makes it very susceptible to resist residues. This not only compromises the spin and charge transport properties, but also the electrical stability of a device due to dielectric charge traps. Here we report the fabrication of CNT spin valves by electron beam lithography using an essentially residue free (on SiO$_2$) low-density polymer that allows the fabrication of optimal polymer mask cross sections without resorting to multiple resist layers. Using this recipe we obtain electrical contacts with a significantly increased electrical stability and yield, for both, thermal and sputter deposition of Py. We present measurements from two devices to discuss the need for an extended data analysis in nanospintronic devices.

\section{Sample fabrication}

Our CNTs are grown by chemical vapor deposition at a temperature of $850^{\circ}$C using methane as source gas and Fe/Ru catalyst nano particles \cite{Li_Dai_JACS_2007}. The substrate is a heavily doped Si wafer acting as a backgate and a $400\,$nm thermal oxide top layer. As shown schematically in Fig.~1a our approach to obtain reproducible magnetic domains and switching characteristics for the ferromagnetic contacts is to fabricate $25\,$nm thick ferromagnetic Permalloy (Py) strips with a large aspect ratio ($\sim100$) \cite{Aurich_Baumgartner_APL97_2010}. These strips are fabricated by electron beam lithography with an electron sensitive polymer resist, followed by metal deposition and a lift-off procedure. To deposit Py we use two techniques: 1) thermal evaporation of Py by an electron gun in a UHV chamber at a base pressure of $\sim10^{-9}\,$mbar, sample cooling to $-30^{\circ}$C and a deposition rate of $\sim0.2\,$\AA/s. 2) DC sputter deposition using an Ar plasma at the power of $35\,$W and an Ar pressure of $\sim6\times10^{-3}\,$mbar in a UHV chamber with a base pressure of $\sim10^{-9}\,$mbar. The fabrication of nanostructures by sputter deposition is often difficult because the sputtered material is scattered at gas particles in the chamber, which leads to a large angular spread that can fill the lithographically defined polymer trench and lead to lift-off problems. We obtain sufficient directionality for the sputter deposition of Py by working with a relatively low Ar pressure and no sample rotation. The sample resides directly above the Py target at a distance of $\sim10\,$cm from the plasma at room temperature. We use a deposition rate of $\sim0.5\,$\AA/s.

\begin{figure*}[b]{
\centering
\includegraphics[width=0.65\linewidth]{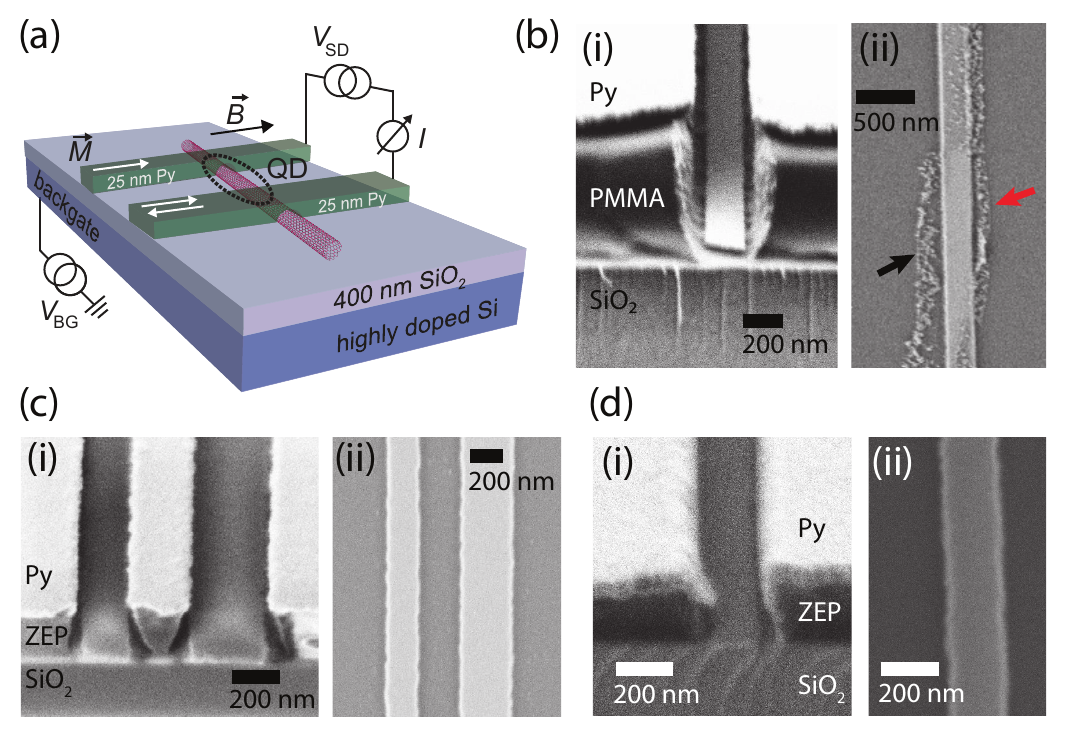}
}
\caption{(Color online) (a) Schematic and measurement scheme of a lateral CNT quantum dot spin valve with Permalloy leads. (b)-(d) Scanning electron microscopy images of (i) cross sections of the metalized strip structures, and (ii) top view of Py strips after the lift-off process. The structures in (b) were obtained with a PMMA mask and $30\,$kV acceleration voltage. The red (bright) arrow points out polymer residues, the black arrow metallic particles. (c) Py strips fabricated using ZEP 520A, $10\,$kV acceleration voltage and thermal evaporation of Py. (d) Strips obtained using ZEP, $10\,$kV acceleration voltage and sputter deposition of Py.}
\end{figure*}

We have systematically investigated the morphology and magnetic properties of Py strips fabricated by electron beam lithography with different resist systems and beam acceleration voltages and identify two fundamental problems rather specific to nanospintronic devices: 1) Py nanoparticles form at the side walls of polymer structures with insufficient under-cuts and are deposited nearby or on the Py strips in the lift-off procedure, 2) resist residues lead to a significant decrease in the yield of obtaining low-ohmic electrical contacts to CNTs (with $<1$M$\Omega$ two-terminal resistance at room temperature).

Both problems are illustrated in scanning electron microscopy (SEM) images in Fig.~1b. Subfigure (i) shows a tilted side view of a $600\,$nm thick poly(methyl methacrylate) (PMMA) mask of a strip after lithography and thermal Py deposition. The Py strip forms at the bottom of the polymer trench. However, due to the large beam acceleration voltage of $30\,$kV used for this structure, the polymer trench is V-shaped with a thin metal film deposited also on the side walls, which often leads to a bad lift-off and large ferromagnetic residues. Subfigure (ii) shows a top view of the resulting Py strip. While the strip appears well defined, we reproducibly find a large number of Py nanoparticles on top and around the strip, as indicated by the black arrow. Such particles can be magnetic with very large characteristic fields, leading to seemingly non-symmetric low-field MR curves.

The polymer profile can be improved significantly by either reducing the SEM acceleration voltage, e.g. to $20\,$kV, or by using an additional more sensitive resist layer, which both lead to a larger under-cut. We have tested the copolymer system PMMA/MA (MA: methacrylic acid) and PMMA(950k)/PMMA(50k) with different PMMA chain lengths. All three methods can be optimised to obtain better undercuts and a significantly reduced number of Py particles on the surface. From this finding we conclude that the particles form at the side walls of V-shaped profiles, which is therefore an essentially geometric effect and independent of the polymer.

The red (bright) arrow in subfigure (ii) of Fig.~1b points out polymer residues, which we identify by the smaller SEM contrast and the much shorter oxygen dry etching times required to remove them in test samples (not shown). We reproducibly find polymer residues for trenches fabricated in the polymer systems PMMA, PMMA/MA \cite{Hagen_diss} and PMMA(950k)/PMMA(50k), with methyl isobutyl ketone (MIBK) and IPA (1:3) as developer and lift-off in warm acetone. This is a well-known issue not specific to nanospintronic devices \cite{Maximov_JVST_2002, Hang_JVST_2002, Macintyre_JVST_2009}. In semiconductor device fabrication, or when contacting metallic parts of the device, the residues can be removed before the next metal deposition by standard cleaning procedures like oxygen plasma etching or Ar sputtering. However, most of these procedures also remove or damage the CNT and graphene parts of a device. We note that also the post-deposition structuring of Py films frequently used for the fabrication of nanometer scaled magnetic devices, e.g. ion milling or chemical etching, also remove carbon with a large rate.

A close to ideal polymer mask cross section and negligible residues can be obtained using the copolymer resist ZEP 520A (ZEP) \cite{ZEP}  and $60\,$s development in n-Amylacetate, stopped in a 9:1 solution of MIBK and IPA, followed by rinsing in IPA. After the metalization, a good lift-off is achieved in a $15\,$min n-methyl-2-pyrrolidone (NMP) bath at $70^{\circ}$C, followed by $30\,$min in acetone at $50^{\circ}$C and rinsing in IPA. We use a $300\,$nm thick ZEP layer, an electron acceleration voltage of $10\,$kV and a typical dose of $\sim34\,\mu$C/cm$^2$, for which we obtain undercuts with a narrow opening at the top of the polymer film, as demonstrated in Fig.~1c (i). This undercut can be tuned systematically by the dose and acceleration voltage. Subfigure (ii) shows a resulting Py strip obtained by thermal Py evaporation. We find no metallic particles or metal flakes and could not detect any resist residues. We obtain similarly clean strips with a slightly increased surface roughness using the ZEP recipe and sputter deposition of Py, as demonstrated in Fig.~1d.

\begin{figure*}[t]{
\centering
\includegraphics[width=0.5\linewidth]{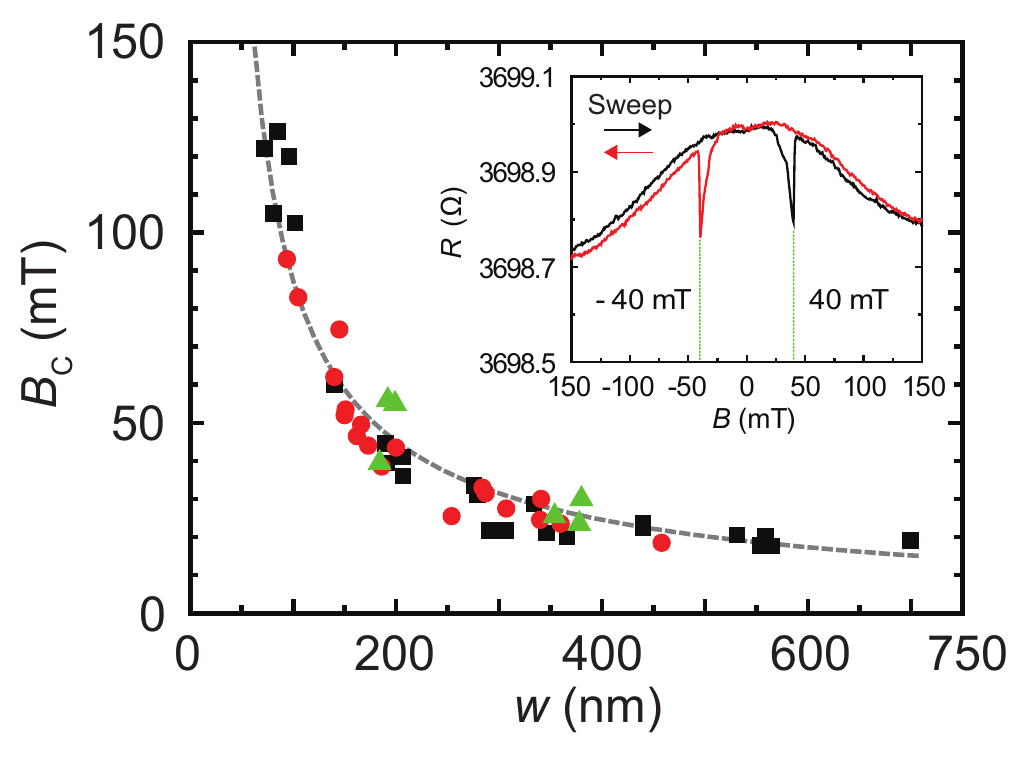}
}
\caption{(Color online) Switching field $B_{c}$ extracted from anisotropic magnetoresistance (AMR) measurements on individual Py strips as a function of the strip width $w$. All strips are $25\,$nm thick and $10\,\mu$m long. The symbols represent values for strips obtained by \textcolor{red}{$\bullet$}~thermal evaporation and ZEP recipe, \textcolor{green}{$\blacktriangle$}~sputter deposition and ZEP masks, and \textcolor{black}{$\blacksquare$}~thermal evaporation and an optimised PMMA/MA recipe. The dashed line is a guide to the eye. The inset shows AMR curves of a $w=180\,$nm sputtered Py strip.}
\end{figure*}

We fabricate long ($10\,\mu$m), thin ($25\,$nm) Py strips with a small width $w$, which forces the magnetisation of the ferromagnetic contacts to lie along the strip axis. The direction can be inverted by an external magnetic field along the axis that switches the magnetisation to the opposite orientation at a characteristic switching field $H_c$ tunable by the width $w$ of the strip \cite{Aurich_Baumgartner_APL97_2010}.
To assess the magnetic properties and material quality of an individual Py strip, it is contacted by Pd contacts to measure the anisotropic magnetoresistance (AMR). An example curve is shown in the inset of Fig.~2, where the resistance $R$ of a $180\,$nm wide strip of sputtered Py is plotted as a function of the external magnetic field $B$ along the strip axis. Sharp characteristic changes in the resistance at $B_{\rm c}\approx \pm 40\,$mT indicate the reversal of the magnetisation \cite{Aurich_Baumgartner_APL97_2010}. The smooth background variation we attribute to a small ($<3^{\circ}$) misalignment between the field and the strip axis, which mixes in the large continuous MR signals obtained when the field is applied perpendicular to the strip axis.

Figure~2 shows the switching fields $B_{\rm c}$ as a function of $w$ for strips obtained by different fabrication techniques. We find that the sputtered and thermally evaporated contacts defined using ZEP exhibit the same dependence on $w$ as the PMMA processed and thermally evaporated Py contacts \cite{Aurich_Baumgartner_APL97_2010}. The switching fields can be distinguished reliably for widths $w<400\,$nm, for which $B_{\rm c}$ increases strongly for smaller $w$.

While the AMR curves of individual Py strips are very reproducible, the resulting MR in a spin valve are more problematic, as will be discussed below. We note already here that AMR experiments are sensitive to the bulk of the material, while in spin valve configurations the last few atomic layers are crucial.


\section{Nanospintronic magnetoresistance experiments}

In this section we demonstrate the need of extended data acquisition and analysis for magnetoresistance devices with non-trivial conductance characteristics.
The magnetoresistance (MR) of a spin valve device is defined in terms of its conductances $G_{\rm p}$ and $G_{\rm ap}$ when the magnetisations in the two contact strips are either parallel (p) or anti-parallel (ap). Similar expressions are easily obtained using the device resistances. Here we define
$$ MR=\frac{G_{\rm p}-G_{\rm a}}{G_{\rm p}+G_{\rm a}},$$
which is more symmetric than the usual definitions and leads to smaller MR values of maximally $100$\%. This definition is more adequate for our purpose because it provides an equal measure for positive and negative MR. In a QD spin valve, the conductance depends on the gate voltage, which tunes the QD level energies. The charging energy and level separation lead to characteristic Coulomb blockade (CB) conductance maxima, with a strongly reduced conductance in between. In the color scale image in Fig.~3 the QD conductance is plotted as a function of the backgate voltage $V_{\rm BG}$ and an increasing magnetic field $B$ (up-sweep) for a QD fabricated with standard PMMA-based lithography. The base temperature in all experiments presented here is $\sim 230\,$mK. The QD conductance has a maximum at $V_{\rm BG}\approx -6.48\,$V and decays rapidly away from this value. A cross section at constant magnetic field is plotted in white. When the magnetic field is increased from negative values beyond $B=0$, a first sharp change ($\Delta B < 1\,$mT) in the conductance pattern occurs at $B_1\approx 20\,$mT, and another at $B_2\approx 30\,$mT. These fields correspond well to the contact switching fields of the two Py strips.

\begin{figure*}[t]{
\centering
\includegraphics[width=0.55\linewidth]{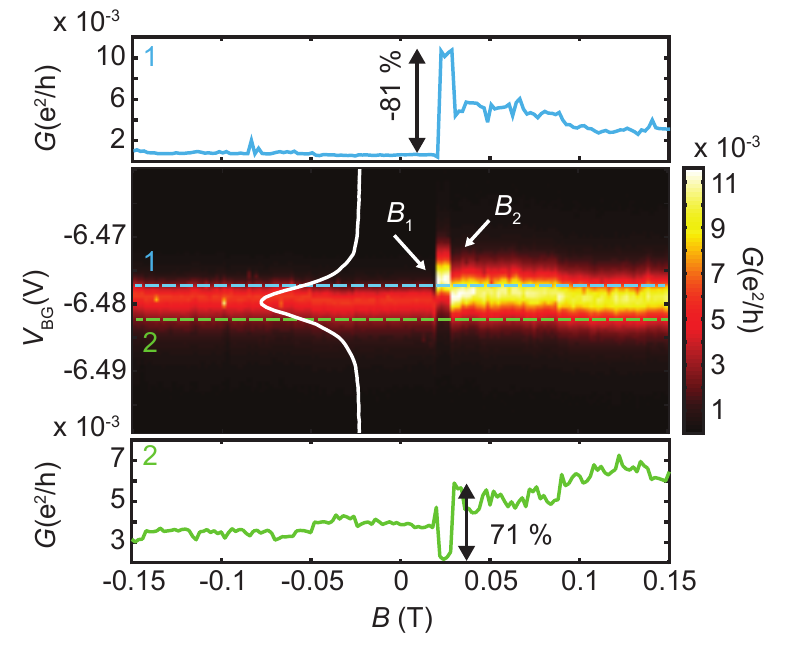}
}
\caption{(Color online) Differential conductance $G$ of a CNT spin valve fabricated by standard PMMA based lithography. $G$ is plotted as a function of the backgate voltage and an external magnetic field applied in parallel to the magnetic contact strips. The top and bottom magnetoresistance (MR) curves are cross sections at the gate voltages indicated by the dashed lines in the main graph.}
\end{figure*}

At $B_1$ the amplitude of the CB resonance increases by a factor of ~$\sim2$ and the peak position shifts by about $\Delta V_{\rm BG}\approx 4\,$mV, which corresponds to an energy shift of $\Delta E\approx 400\,\mu$eV or to almost the resonance width.
While the amplitude of the CB resonance increases by almost a factor of 2 at $B_1$ starting at the low field side, it does not change at $B_2$ and is reduced slightly only at higher fields.
At $B_2$ the resonance position switches back roughly to the same gate position as for $B<B1$. In a standard magnetoresistance (MR) measurement the conductance is recorded as a function of $B$ alone, which corresponds to cross sections in Fig.~3 at a fixed gate voltage. Two examples for slightly off-resonance voltages are shown on top and below the main figure: at a more negative gate voltage (green dashed line, 2) we find a decrease in conductance for the anti-parallel magnetisations, $B_1<B<B_2$, which corresponds to an increased resistance and a positive MR of $\sim 70$\%. Off-resonance for a more positive gate voltage (blue dashed line, 1) the MR at fixed voltage is negative, $MR\approx -80$\%. These large values are almost exclusively due to the large shift of the resonance position.
In the simplest model by Julli\`ere for tunneling MR \cite{Julliere_PLA54_1975} one would expect $MR=P_1P_2 \approx9-25$\% when using $P_1=P_2=0.3-0.5$ for the tunneling polarizations in the two F-contacts \cite{polarization_clarification}. These values rather correspond to the amplitude modulation ($MR<30$\%) than to the MR observed in cross sections.

We will discuss shifts of the conductance features in the MR in more detail in the next section and only point out that while the MR at $B_1$ and $B_2$ might be described by a simple spin valve model, the increase of $G$ for $B>B_2$ with respect to $B<B_1$ is more difficult to explain since it suggests a difference between the two parallel configurations, a phenomenon possibly related to the single switching behavior reported before \cite{Lee_Nanotech24_2013}.

\section{Negative magnetoresistance over complete orbital}

The electrical stability and reproducibility of the QD spin valve signals is considerably improved for devices fabricated using the ZEP recipe introduced above. We analyze in more detail the data shown in Fig.~4 measured on a sample with sputtered Py contacts.

In Fig.~4a the QD spin valve conductance $G$ is plotted for a large backgate voltage interval at a base temperature of $\sim230\,$mK. The CB peaks occur in groups of four consistent with the spin and valley degeneracy of a CNT orbital. Such a pattern suggests that the CNT segment forming the QD is relatively clean \cite{Jung_Baumgartner_Nanoletters_2013}. From charge stability diagrams (not shown) we find the lever arm of the backgate to the QD $\alpha_{\rm BG}\approx0.14$, a charging energy of $\sim4.5\,$meV and a level spacing of $\sim3.5\,$meV. We estimate the source, drain and backgate capacitances as $C_{\rm S}\approx 23.6\,$aF, $C_{\rm D}\approx 6.3\,$aF and $C_{\rm BG}\approx 4.9\,$aF. From the CB maxima of $\sim0.5\frac{e^2}{h}$, the average broadening of the peaks $\sim 2.4\,$meV and using the Breit-Wigner form for resonant tunneling at low temperatures ($kT<<\Gamma$), we find for the tunnel couplings of the QD to source and drain $\Gamma_{\rm S}\approx 2.0\,$meV and $\Gamma_{\rm D}\approx0.4\,$meV, which gives a relatively small coupling asymmetry of $\Gamma_{\rm S}/\Gamma_{\rm D}\approx5$ (we chose the larger value as $\Gamma_{\rm S}$).

We now focus on the four CB peaks highlighted in Fig.~4a by the red rectangle, which originate from the same four-fold degenerate QD orbital. In Fig.~4b and 4c the QD conductance $G$ is plotted for this gate voltage interval and as a function of an external magnetic field $B$ along the Py contact strips. In Fig.~4b the field is increased from negative values, while in 4c it is decreased, starting from positive values. The magnetisations were saturated at $\pm150\,$mT before the respective sweep. In the up sweep in Fig.~4b, $G$ is larger for $27\leq B \leq 33\,$mT, which usually is identified as the anti-parallel configuration of the contact magnetisations.
In the down sweep the magnetisation switching occurs at negative fields and we find an increased conductance for $-32\leq B \leq -22\,$mT. The variation between the absolute values of the switching fields in the up and down sweeps are compatible with the variation observed in the corresponding AMR curves.

\begin{figure*}[t]{
\centering
\includegraphics[width=0.9\linewidth]{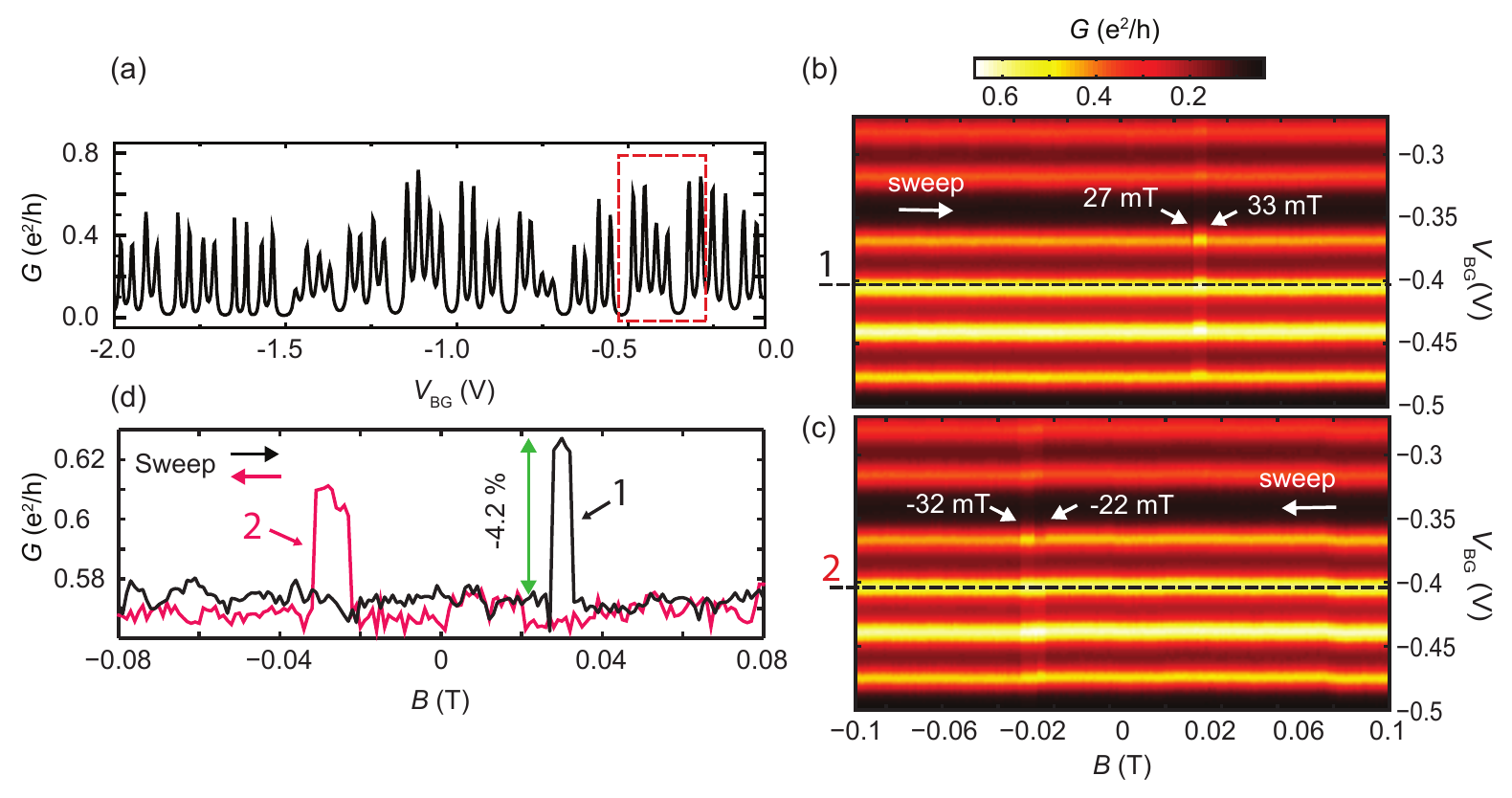}
}
\caption{(Color online) (a) differential conductance $G$ of a CNT spin valve for a large backgate voltage interval. (b), (c) $G$ as a function of the backgate voltage $V_{\rm BG}$ and the external magnetic field $B$ for increasing (b) and decreasing field (c). (d) Up and down sweeps at fixed $V_{\rm BG}$, indicated as dashed lines in (b) and (c).}
\end{figure*}

In Fig.~4d shows the up and down sweeps at a fixed backgate voltage, indicated by the dashed lines in Figs.~4b and 4c. We find a sharp switching of the conductance at the Py strip switching fields, which corresponds to a MR of $\sim-4$\%. The MR is negative for all gate voltages, which we now discuss in more detail.

In devices with a variable conductance, e.g. as a function of the backgate voltage, the origin of the MR signal can lie in changes of the width, position and amplitude of the conductance feature. In Fig.~5a we plot the CB oscillations indicated in Fig.~4a as a function of $V_{\rm BG}$ for the different magnetisation configurations. The two parallel configurations lead to identical conductances, which demonstrates the reproducibility of both, the magnetic and electronic structures in the device. The anti-parallel configuration, however, deviates significantly from the parallel. The resulting MR vs $V_{\rm BG}$ curve is plotted in Fig.~5b (full red line). The MR is negative for almost all backgate voltages and shows a MR modulation of $\sim10$\% on an offset of about $-5$\%. The modulation is correlated with the gradient $dG/dV_{\rm BG}$ of $G$, i.e. it is largest at the gate voltages where $G$ has the largest slopes, which suggests that the MR is caused mainly by a shift of the CB resonances.

In the next step we fit the data with multiple Lorenzians to extract the amplitude, width and position of the individual CB peaks (no background is subtracted). The resulting parameters are plotted in Figs.~5c-e for the third CB peak highlighted by an asterisk in Fig.~5a. Compared to the parallel magnetisation configurations, the anti-parallel shows an increase in amplitude and width by $\sim 4$\% and $4.5$\%, respectively, and a shift of $\sim1.0\,$mV, which corresponds to $\sim140\,\mu$eV or $\sim6$\% of the peak width. We obtain similar values for the other CB peaks. All peaks are shifted by the same absolute value within experimental errors.

\begin{figure*}[t]{
\centering
\includegraphics[width=0.6\linewidth]{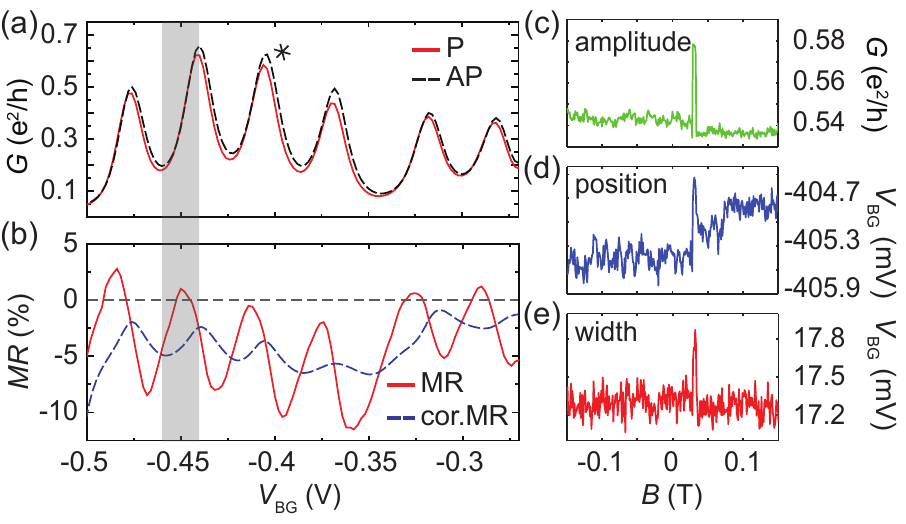}
}
\caption{(Color online) (a) Conductance $G$ as a function of the backgate voltage $V_{\rm BG}$ for the magnetisation configurations (both parallel $(\uparrow,\uparrow)$, $(\downarrow, \downarrow)$, and antiparallel $(\uparrow,\downarrow)$, $(\downarrow, \uparrow)$. The corresponding MR curves are shown in (b). (c)-(e) Amplitude, position and width of the third CB resonance peak (asterisk) as a function of the applied magnetic field $B$, extracted from the fits described in the text.}
\end{figure*}

The extracted parameters allow us to investigate the respective impact on the MR, for example by calculating the MR from the measured curve for the parallel magnetisations and a shifted curve in the anti-parallel case. The result is plotted in Fig.~5b as blue dashed line (cor. MR) and has MR maxima at gate voltages where also $G$ has maxima, as expected if the shifts were corrected precisely enough. The MR variation on this curve is only $\sim3$\% with a slightly smaller negative offset than in the original data.

\section{Discussion}

We generally find a better electrical stability and larger contact yield for QDs fabricated with the presented recipes. We note, however, that also with this method we obtain samples that show gate-dependent or temporal charge rearrangements, which we tentatively attribute to surface impurities on the substrate. Nevertheless, these methods lead to devices with reproducible MR with a yield of $\sim40$\%, a clear improvement compared to previous methods that yielded useful devices only rarely ($<5$\%).

A periodic modulation of the MR with the CB oscillations was observed already earlier and modeled by spin dependent effective tunnel rates \cite{Sahoo_Kontos_Schoenenberger_NaturePhys1_2005}. In this simple model one can construct negative MR signals for a strongly asymmetric QD coupling to the contacts, which can lead to a negative offset for strongly overlapping resonances. The change of the effective tunnel couplings at the switching fields could in principle also result in a change of the resonance widths. However, this model requires strongly asymmetric tunnel couplings and predicts that the MR maxima occur near the conductance minima, both in contrast to our observations.  This model does not produce shifts in the CB resonance energies, either.

More elaborate models \cite{Cottet_Schoenenberger_SemicondSciTechnol21_2006,  Koller_Grifoni_Paaske_PRB85_2012} predict major contributions to the MR from shifts of the CB resonances in an effective magnetic field, caused either by spin dependent electron scattering at the QD-contact interfaces or by a spin-dependent renormalization of the QD energy levels. Characteristic for both mechanisms is that the sign of the shifts depends on the spin state of the CB resonance. Specifically, of the four states in a CNT orbital, two should be shifted in energy opposite to the other two. None of the models predict identical shifts for all four peaks, nor a negative offset of the MR.

Another mechanism that results in a constant shift in the anti-parallel configuration is the magneto-Coulomb effect (MCE) \cite{Shimada_PRB64_2001, van_der_Molen_vanWees_PRB73_2006}. The opposite Zeemann shifts and the different density of states at the Fermi energy of the majority and minority bands leads to a rearrangement of electrons between the bands, which is compensated by a change in the electrical potential. We estimate an MCE shift of the QD resonances in an external magnetic field $B$ of $\Delta V_{\rm BG}/B=\frac{1}{2e}\frac{C_{\rm S}+C_{\rm D}}{C_{\rm BG}}Pg\mu_{\rm B}\approx 300\,\mu$V/T. In the last step we used $P=0.8$ as an upper limit of the (thermodynamic) Py polarization in both leads, the Land\'{e} $g$-factor in thick ($>15\,$nm) Py films of $g= 2.1$ \cite{Nibarger_APL83_2003} and the Bohr Magneton $\mu_{\rm B}$. With the same parameters one obtains a total change in position of $\Delta V_{\rm BG}\approx15\,\mu$V when sweeping the field beyond both switching fields. The negligible slope observed for the peak positions is consistent with the small value obtained in these estimates. However, the same parameters also predict a negligible change at the switching fields. We note that also the qualitative curve shape observed in the experiments does not follow the triangular characteristics at the switching fields of the MCE \cite{van_der_Molen_vanWees_PRB73_2006}.

A quite natural explanation of our experimental findings is that one Py strip does not couple directly to the QD, but rather to an anti-ferromagnetically coupled contact area. Such effects could occur at the chemical bonds between the metal/CNT interface \cite{DeTeresa_Science286_1999}, or in oxidized layers of the  magnetic material that are strongly coupled to the bulk. The latter coupling could depend on the thickness of the oxide and explain why both contacts are not coupled identically to the CNT, a phenomenon well known from non-magnetic metal contacts to CNTs.
This scenario explains the sign reversal of the MR gate modulation and offset, but not the peak shifts at the switching fields. A physically more intriguing scenario would be a spatially varying spin susceptibility (RKKY interaction) in the CNT segments connecting the QD \cite{Stano_Klinovaja_Yacobi_Loss_PRB88_2013}, which could lead to a rotation of the injected spins depending on the distance from the ferromagnetic contact - an effect that can persist over large length scales due to the low dimension of the CNT and the strong electron-electron interactions.

\section{Conclusions}
We report a recipe for essentially residue free electron beam lithography, useful to improve the fabrication of carbon based nanospintronic devices. We obtain very reproducible magnetic and electrical contact properties for sputtered and thermally deposited Permalloy when both processes are optimised individually. Using these recipes, we obtain an improved yield of electrical contacts and more reproducible features in the MR of carbon nanotube quantum dot spin valves.
Since the MR in nanospintronic devices not only depends on the magnetic orientation of the contacts, but also on the electrostatic environment (e.g. gates), it is necessary to expand the standard magnetic field sweeps to three-dimensional maps that also contain a variable gate voltage to track the origin of the observed MR. We demonstrate this idea with two devices and show that the major contribution to the MR can be due to a shift in the conductance features. From the discussion of several mechanisms specific to nanospintronic devices we tentatively conclude that interface properties might be crucial to explain the presented magnetoresistance characteristics.

\ack
We gratefully acknowledge useful discussions with B. Hickey and M. Elkin. This work is financially supported by the Swiss National Science Foundation (SNF), the Swiss NCCR QSIT and NCCR Nano, the ERC project QUEST and the FP7 project SE2ND.


\begin{thebibliography}{10}
\bibitem{polarization_clarification}
Here we use the so-called tunneling polarization that also comprises the tunneling matrix elements and not only the density of states.
\end{thebibliography}
\end{document}